\documentclass[reprint,amsmath,amssymb,aps,]{revtex4-2}
\usepackage{bm}
\usepackage{dcolumn}
\usepackage{lineno}
\usepackage{graphicx}
\usepackage{subfigure}
\usepackage{epstopdf}
\usepackage{float}
\usepackage{threeparttable}
\usepackage[colorlinks, linkcolor=blue, anchorcolor=blue, citecolor=blue, urlcolor=black]{hyperref}
\usepackage{cmap} 
\usepackage{mathtext}
\usepackage{mathrsfs}

\newcommand{\nJpsi}{1.0087} \newcommand{\nJpsiErr}{0.0044} \newcommand{\nJpsiIncMC}{1.0011\times10^{10}}
\newcommand{\Fconv}{1.010} \newcommand{\FconvEsta}{0.004} \newcommand{\FconvEsys}{0.005} \newcommand{\FconvEPercent}{0.64}
\newcommand{\brInc}{1.067} \newcommand{\brIncEsta}{0.005} \newcommand{\brIncEsys}{0.023}
\newcommand{\brCa}{39.86} \newcommand{\brCaEsta}{0.04} \newcommand{\brCaEsys}{0.99} \newcommand{\reSysCa}{2.49}
\newcommand{\brCb}{31.96} \newcommand{\brCbEsta}{0.07} \newcommand{\brCbEsys}{0.84} \newcommand{\reSysCb}{2.62}
\newcommand{\brCc}{4.38} \newcommand{\brCcEsta}{0.02} \newcommand{\brCcEsys}{0.10} \newcommand{\reSysCc}{2.39}
\newcommand{\brCd}{23.04} \newcommand{\brCdEsta}{0.03} \newcommand{\brCdEsys}{0.54} \newcommand{\reSysCd}{2.35}
\newcommand{\effInc}{8.090} \newcommand{\effIncErr}{0.009} \newcommand{\effCa}{48.46} \newcommand{\effCaErr}{0.01} \newcommand{\effCb}{8.230} \newcommand{\effCbErr}{0.004} \newcommand{\effCc}{42.86} \newcommand{\effCcErr}{0.01} \newcommand{\effCd}{24.73} \newcommand{\effCdErr}{0.01}
\newcommand{\raCb}{0.802} \newcommand{\raCbEsta}{0.002} \newcommand{\raCbEsys}{0.014}
\newcommand{\raCc}{0.110} \newcommand{\raCcEsta}{0.001} \newcommand{\raCcEsys}{0.002}
\newcommand{\raCd}{0.578} \newcommand{\raCdEsta}{0.001} \newcommand{\raCdEsys}{0.008}
\newcommand{\sysIncExSys}{2.20} \newcommand{\sysIncExSysSta}{2.24}
\newcommand{\sysTrkOneCc}{0.14} \newcommand{\sysTrkOneCd}{0.13}
\newcommand{\sysTrkCc}{0.28} \newcommand{\sysTrkCd}{0.26}
\newcommand{\sysGamOneCa}{0.21} \newcommand{\sysGamOneCb}{0.18} \newcommand{\sysGamOneCc}{0.17} \newcommand{\sysGamOneCd}{0.18}
\newcommand{\sysGamCa}{0.64} \newcommand{\sysGamCb}{1.26} \newcommand{\sysGamCc}{0.34} \newcommand{\sysGamCd}{0.53}
\newcommand{\sysFitRInc}{0.87} \newcommand{\sysFitRCa}{0.08} \newcommand{\sysFitRCb}{0.11} \newcommand{\sysFitRCc}{0.32} \newcommand{\sysFitRCd}{0.08}
\newcommand{\sysBkgInc}{1.09}      
\newcommand{\sysKmfitCa}{0.11} \newcommand{\sysKmfitCb}{0.29} \newcommand{\sysKmfitCc}{0.22} \newcommand{\sysKmfitCd}{0.09}

\begin{document}
\title{\boldmath Measurement of the absolute branching fractions of $J/\psi\to\gamma\eta$ and $\eta$ decay modes}
\author{
\begin{small}
\begin{center}
M.~Ablikim$^{1}$, M.~N.~Achasov$^{10,b}$, P.~Adlarson$^{67}$, S.~Ahmed$^{15}$, M.~Albrecht$^{4}$, R.~Aliberti$^{28}$, A.~Amoroso$^{66A,66C}$, M.~R.~An$^{32}$, Q.~An$^{63,49}$, X.~H.~Bai$^{57}$
% ...
\\
\vspace{0.2cm}
(BESIII Collaboration)\\
\vspace{0.2cm} {\it
$^{1}$ Institute of High Energy Physics, Beijing 100049, People's Republic of China\\
% ...
$^{4}$ Bochum Ruhr-University, D-44780 Bochum, Germany\\
% ...
$^{10}$ G.I. Budker Institute of Nuclear Physics SB RAS (BINP), Novosibirsk 630090, Russia\\
% ...
$^{15}$ Helmholtz Institute Mainz, Staudinger Weg 18, D-55099 Mainz, Germany\\
% ...
$^{28}$ Johannes Gutenberg University of Mainz, Johann-Joachim-Becher-Weg 45, D-55099 Mainz, Germany\\
% ...
$^{32}$ Liaoning Normal University, Dalian 116029, People's Republic of China\\
% ...
$^{49}$ State Key Laboratory of Particle Detection and Electronics, Beijing 100049, Hefei 230026, People's Republic of China\\
% ...
$^{57}$ University of Jinan, Jinan 250022, People's Republic of China\\
% ...
$^{63}$ University of Science and Technology of China, Hefei 230026, People's Republic of China\\
% ...
$^{66}$ University of Turin and INFN, (A)University of Turin, I-10125, Turin, Italy; (B)University of Eastern Piedmont, I-15121, Alessandria, Italy; (C)INFN, I-10125, Turin, Italy\\
$^{67}$ Uppsala University, Box 516, SE-75120 Uppsala, Sweden\\
% ...
$^{b}$ Also at the Novosibirsk State University, Novosibirsk, 630090, Russia\\
% ...
}\end{center}
\vspace{0.4cm}
\end{small}
}

\date{\today}


	\begin{abstract}
		Based on a data sample of $(\nJpsi\pm\nJpsiErr)\times10^{10}$ $J/\psi$ events collected by the BESIII detector at the BEPCII accelerator, the absolute branching fraction (BF) of the decay $J/\psi\to\gamma\eta$ is measured with high precision using events in which the radiative photon converts to $e^+e^-$. Using the measured absolute BF of $J/\psi\to\gamma\eta$, the absolute BFs of four dominant $\eta$ decay modes are measured for the first time. The results are $\mathcal{B}(J/\psi\to\gamma\eta)=(\brInc\pm\brIncEsta\pm\brIncEsys)\times 10^{-3}$, $\mathcal{B}(\eta\to\gamma\gamma)=(\brCa\pm\brCaEsta\pm\brCaEsys)\%$, $\mathcal{B}(\eta\to\pi^0\pi^0\pi^0)=(\brCb\pm\brCbEsta\pm\brCbEsys)\%$, $\mathcal{B}(\eta\to\pi^+\pi^-\pi^0)=(\brCd\pm\brCdEsta\pm\brCdEsys)\%$, and $\mathcal{B}(\eta\to\pi^+\pi^-\gamma)=(\brCc\pm\brCcEsta\pm\brCcEsys)\%$, where the first and second uncertainties are statistical and systematic, respectively. The results are consistent with the world average values
within two standard deviations.
	\end{abstract}
	\maketitle

	\section{Introduction}\label{sec:introduction}\vspace{-0.3cm}
		As two members of the ground-state nonet of pseudo-scalar mesons, the $\eta$ and $\eta'$ mesons play an important part in understanding low energy Quantum Chromodynamics (QCD)~\cite{etaRev2007,etaRev2019}. Precise measurements of their branching fractions (BFs) are important for a wide variety of physics topics. For example, the decay widths of $\eta,~\eta'\to\gamma\gamma$ are related to the quark content of the two mesons~\cite{IntroTwoGam}, the BFs of $\eta,~\eta'\to 3\pi$ decays can provide valuable information on light quark masses~\cite{IntroQuarkMass}, the BFs of $\eta,~\eta'\to\pi^+\pi^-\gamma$ decays are related to details of chiral dynamics~\cite{IntroChiralDynamic1,IntroChiralDynamic2}, and the BFs of some rare decays of the $\eta$ and $\eta'$ can test fundamental QCD symmetries~\cite{IntroQCDSymmetry} and probe for physics beyond the standard model~\cite{IntroBSM}. As the BFs of the rare decays are obtained via normalization to the dominant decay modes, a precise determination of the BFs of the dominant decay modes of the $\eta$ and $\eta'$ is essential. While the absolute BFs of dominant $\eta'$ decays have been measured with high precision by the BESIII experiment~\cite{bibGamEtaP}, no absolute BFs of $\eta$ decays have yet been measured. The exclusive BFs of the $\eta$ summarized by the Particle Data Group (PDG)~\cite{pdg2019} are all relative measurements. This is due to the difficulty of tagging inclusive decays of the $\eta$. The most precise measurements so far are from the CLEO experiment~\cite{cleoEta}, where the BFs were presented under the assumption that the five dominant decay modes measured in their work account for 99.9\% of all $\eta$ decays.

		In the previous work by BESIII on $\eta'$ decays~\cite{bibGamEtaP}, absolute BFs were measured using a specially developed method which allows tagging inclusive decays of the $\eta'$. In this work, using a similar but optimized method and a much larger $J/\psi$ sample, inclusive decays of the $\eta$ are tagged and the absolute BFs of dominant $\eta$ decay modes are measured for the first time.

	\section{BESIII Detector}
	The BESIII detector~\cite{Ablikim:2009aa} records symmetric $e^+e^-$ collisions provided by the BEPCII storage ring~\cite{Yu:IPAC2016-TUYA01}, which operates with a design luminosity of $1\times10^{33}$~cm$^{-2}$s$^{-1}$ in the center-of-mass energy range from 2.0 to 4.9 GeV. BESIII has collected large data samples in this energy region~\cite{Ablikim:2019hff}. The cylindrical core of the BESIII detector covers 93\% of the full solid angle and consists of a beam pipe, a helium-based multilayer drift chamber~(MDC), a plastic scintillator time-of-flight system~(TOF), and a CsI(Tl) electromagnetic calorimeter~(EMC), which are all enclosed in a superconducting solenoidal magnet providing a 1.0~T (0.9~T in 2012) magnetic field. Around 10.8\% of $J/\psi$ events were collected in 2012. The beam pipe has two layers with 2 mm gaps between them. The inner layer diameter is 63 mm with a thickness of 0.8 mm, while the thickness of the outer layer is 0.5 mm. The inner diameter of the MDC is 118 mm. The solenoid is supported by an octagonal flux-return yoke with resistive plate counter muon identification modules interleaved with steel. The charged-particle momentum resolution at $1~{\rm GeV}/c$ is $0.5\%$, and the $dE/dx$ resolution is $6\%$ for electrons from Bhabha scattering. The EMC measures photon energies with a resolution of $2.5\%$ ($5\%$) at $1$~GeV in the barrel (end cap) region. The time resolution in the TOF barrel region is 68~ps, while that in the end cap region is 110~ps. The end cap TOF system was upgraded in 2015 using multi-gap resistive plate chamber technology, providing a time resolution of 60~ps~\cite{etof}.

	\section{Dataset and MC Simulation}\label{sec_dataset}

	A sample of $(\nJpsi\pm\nJpsiErr)\times10^{10}$ $J/\psi$ events collected by BESIII is used for this analysis. The total number of $J/\psi$ events collected in the years of 2009, 2012, 2018 and 2019 at BESIII is determined using inclusive $J/\psi$ decays with the method described in Ref.~\cite{nJpsi0912}. For the selected inclusive $J/\psi$ events, the background due to QED processes, beam-gas interactions, and cosmic rays is estimated using the continuum data samples at $\sqrt{s} = 3.08$ GeV. The detection efficiency for the inclusive $J/\psi$ decays is obtained using the data sample of $\psi(3686) \rightarrow \pi^+\pi^-J/\psi$. The efficiency difference between the $J/\psi$ produced at rest and the $J/\psi$ from the decay $\psi(3686)\rightarrow\pi^+\pi^-J/\psi$ is estimated by comparing the corresponding efficiencies in MC simulation. The uncertainties related to the signal MC model, track reconstruction efficiency, fit to the $J/\psi$ mass peak, background estimation, noise mixing, and reconstruction efficiency for the pions recoiling against the $J/\psi$ are studied. Finally, the number of $J/\psi$ events collected at BESIII is determined to be $N_{J/\psi}= (10087\pm44)\times10^{6}$.

	Simulated data samples are produced with a {\sc geant4}-based~\cite{geant4} Monte Carlo (MC) package~\cite{MCPackage}, which includes the geometric description of the BESIII detector and the detector response~\cite{BESIIIDetectorShape,BESIIIDetectorShapeB}. They are used to determine the detection efficiency and estimate the backgrounds. The simulation includes the beam energy spread and initial state radiation (ISR) in the $e^+e^-$ annihilations modeled with the generator {\sc kkmc}~\cite{kkmc}. A sample of $\nJpsiIncMC$ simulated inclusive $J/\psi$ events is used to estimate the background events. This inclusive MC sample includes both the production of the $J/\psi$ resonance and the continuum processes incorporated in {\sc kkmc}. The known decay modes are modeled with {\sc evtgen}~\cite{evtgen} using BFs taken from the PDG~\cite{pdg2019}, and the remaining unknown charmonium decays are modeled with {\sc lundcharm}~\cite{lundcharm}. Final state radiation (FSR) from charged final state particles is incorporated using {\sc photos}~\cite{photos}.

	In addition, a sample of $1\times 10^8$ $J/\psi\to\gamma\eta$ simulated events is generated to determine the detection efficiency. In the simulation, the $\eta$ decay BFs from the PDG~\cite{pdg2019} are used, and the decay modes are described with theoretical models that have been validated in previous works, as listed in Table~\ref{table_generators}. To study background distributions, exclusive MC samples for specific background processes, such as $e^+e^-\to\gamma\gamma$, are generated. The simulated processes and the corresponding theoretical models are listed in Table~\ref{table_generators}.

		\begin{table}[htbp]
		\centering
		\caption{Generator models used for MC simulations.}\label{table_generators}
		\begin{tabular}{lc}\hline\hline
			Decay mode & Generator model\\\hline
			$J/\psi\to\gamma\eta$ & Helicity amplitude~\cite{evtgenGuide}\\
			$\eta\to\gamma\gamma$ & Phase space~\cite{evtgenGuide}\\
			$\eta\to\pi^0\pi^0\pi^0$ & Dalitz plot analyses~\cite{etaTo3PiGen}\\
			$\eta\to\pi^+\pi^-\pi^0$ & Dalitz plot analyses~\cite{etaTo3PiGen}\\
			$\eta\to\pi^+\pi^-\gamma$ & Box anomaly proceed~\cite{etaToGamLLGen}\\
			$\eta\to\gamma e^+e^-$ & Electromagnetic Dalitz decays~\cite{etaToGamLLGen}\\
			$\eta\to\gamma\mu^+\mu^-$ & Electromagnetic Dalitz decays~\cite{etaToGamLLGen}\\
			Other $\eta$ decays & Phase space~\cite{evtgenGuide}\\
			$e^+e^-\to\gamma\gamma$ & BABAYAGA~\cite{babayaga1,babayaga2,babayaga3}\\
			$e^+e^-\to e^+e^-$ & BABAYAGA~\cite{babayaga1,babayaga2,babayaga3}\\
			$J/\psi\to e^+e^-\eta$ & Electromagnetic Dalitz decays~\cite{eeEtaGen}\\
			$J/\psi\to\pi^+\pi^-\pi^0$ & Dalitz plot analyses~\cite{rhoPiGen}\\
			$J/\psi\to\omega\eta$ & Helicity amplitude~\cite{evtgenGuide}\\
			$J/\psi\to\omega\pi^0$ & Helicity amplitude~\cite{evtgenGuide}\\
			$J/\psi\to\gamma\eta'$ & Helicity amplitude~\cite{evtgenGuide}\\
			$\eta'$ decays & Same as in~\cite{bibGamEtaP}\\
			\hline\hline
		\end{tabular}
		\end{table}

	\section{Event Selection and Background Analysis}
		To tag inclusive decays of the $\eta$, $J/\psi\to\gamma\eta$ events in which the radiative photon converts to an $e^+e^-$ pair are selected using the Photon Conversion Finder (PCF) package~\cite{PCF}. Reconstructed photon conversion events have an energy resolution twice as good as photons reconstructed in the EMC. The signal of the $\eta$ meson is extracted from the recoil mass spectrum of the $e^+e^-$ conversion pair, $M_{\rm{recoil}}(e^+e^-)$. The BF of $J/\psi\to\gamma\eta$ is calculated with
		\begin{equation}
			\mathcal{B}(J/\psi\to\gamma\eta) = \frac{{{N_{\gamma\eta}^{\rm{obs}}}}}{N_{J/\psi}\cdot\varepsilon_{\gamma\eta}\cdot f}.\label{eq:gamEtaBF}
		\end{equation}
		Here, $N_{\gamma\eta}^{\rm{obs}}$ is the number of observed $J/\psi\to\gamma\eta$, $\gamma\to e^+e^-$ events, $N_{J/\psi}$ is the total number of $J/\psi$ decays, $\varepsilon_{\gamma\eta}$ is the detection efficiency obtained from MC simulation, and $f$ is a factor used to correct for the difference in photon conversion efficiencies between data and MC simulation.

		After that, $J/\psi\to\gamma\eta,~\eta\to X$ events are reconstructed to study the $\eta$ decay BFs. Here, $X$ stands for one of the four dominant $\eta$ decay modes: $\gamma\gamma$, $\pi^0\pi^0\pi^0$, $\pi^+\pi^-\pi^0$, and $\pi^+\pi^-\gamma$. To improve statistics, the radiative photons are required to be detected in the EMC istead of converting to $e^+e^-$. The absolute BFs of $\eta\to X$ are then obtained with 
		\begin{equation}
			\mathcal{B}(\eta\!\to\! X) \!=\! \frac{ N_{X}^{\rm{obs}} }{\varepsilon_{X}\!\cdot\! N_{J/\psi}\!\cdot\!\mathcal{B}(J/\psi\!\to\!\gamma\eta)} \!=\! \frac{ N_{X}^{\rm{obs}} }{ \varepsilon_{X} } \cdot \frac{\varepsilon_{\gamma\eta}\!\cdot\! f}{N_{\gamma\eta}^{\rm{obs}}}.\label{eq:etaBF}
		\end{equation}
Here $N_{X}^{\rm{obs}}$ denotes the number of observed $J/\psi\to\gamma\eta$, $\eta\to X$ events, and $\varepsilon_{X}$ the MC-determined reconstruction efficiency.

	\subsection{Inclusive channel}\label{sec:inclusive_selections}
		To select $J/\psi\to\gamma\eta$ events where the radiative photon converts to $e^+e^-$, candidate events are required to have at least two oppositely charged tracks. The charged tracks are reconstructed using information from the MDC and are required to pass within $\pm30$ cm of the run-by-run determined interaction point (IP) along the beam direction. They must also have a polar angle ($\theta$) within the range $|\!\cos\theta|<0.93$, where $\theta$ is defined with respect to the MDC axis. In the next step, particle identification (PID) requirements are applied. The combined information from the specific energy loss in the MDC ($dE/dx$), TOF, and EMC is used to calculate the probability that the track originates from an electron or positron. This probability is then compared to the corresponding probability that the track originates from a muon, pion, kaon, or proton. A track is assumed to be an electron or positron if its probability is larger than the other particle hypotheses. The event is kept for further analysis if there is at least one positron and one electron candidate.

\begin{figure}[htbp]
		\begin{center}
			\subfigure{\includegraphics[width=0.41\columnwidth]{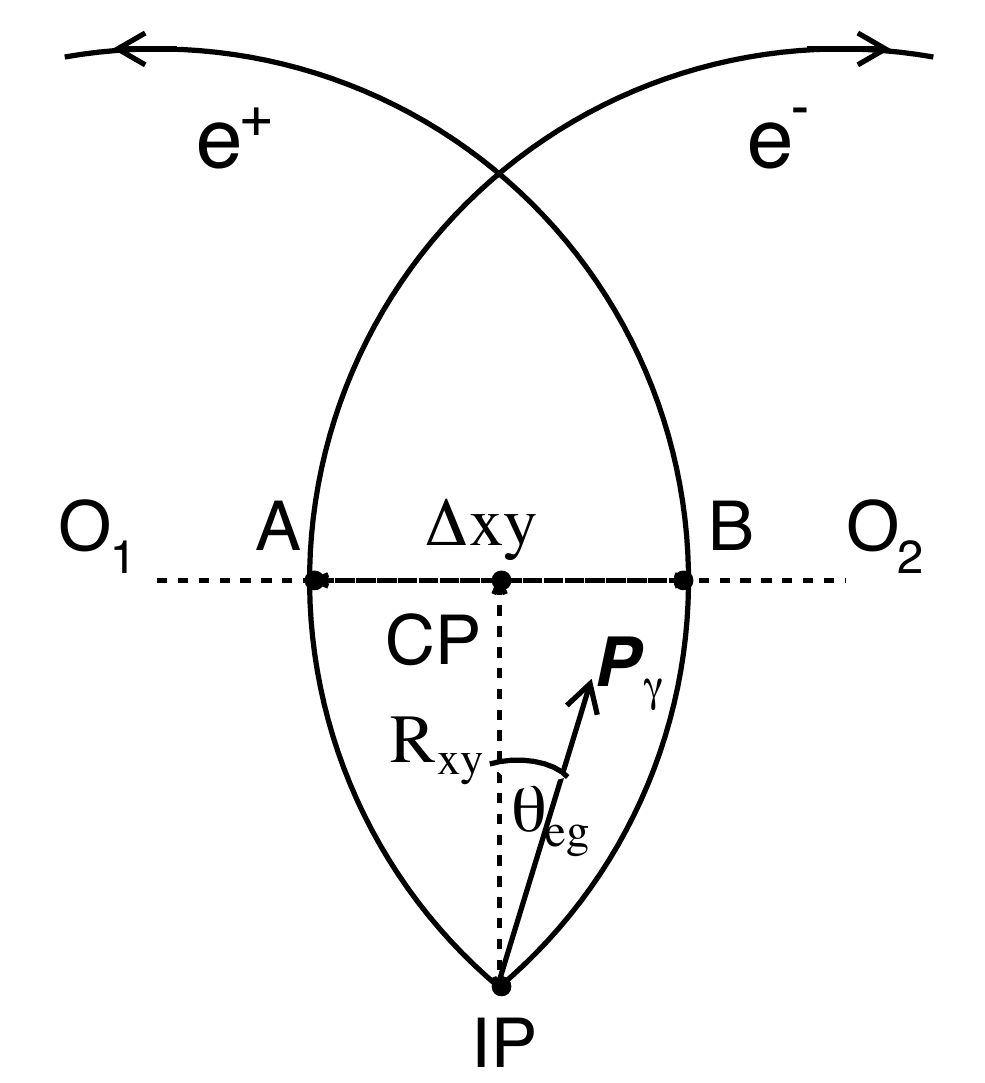}\label{fig_gamConvA}}
			\put(-90,10){(a)}
			\subfigure{\includegraphics[width=0.57\columnwidth]{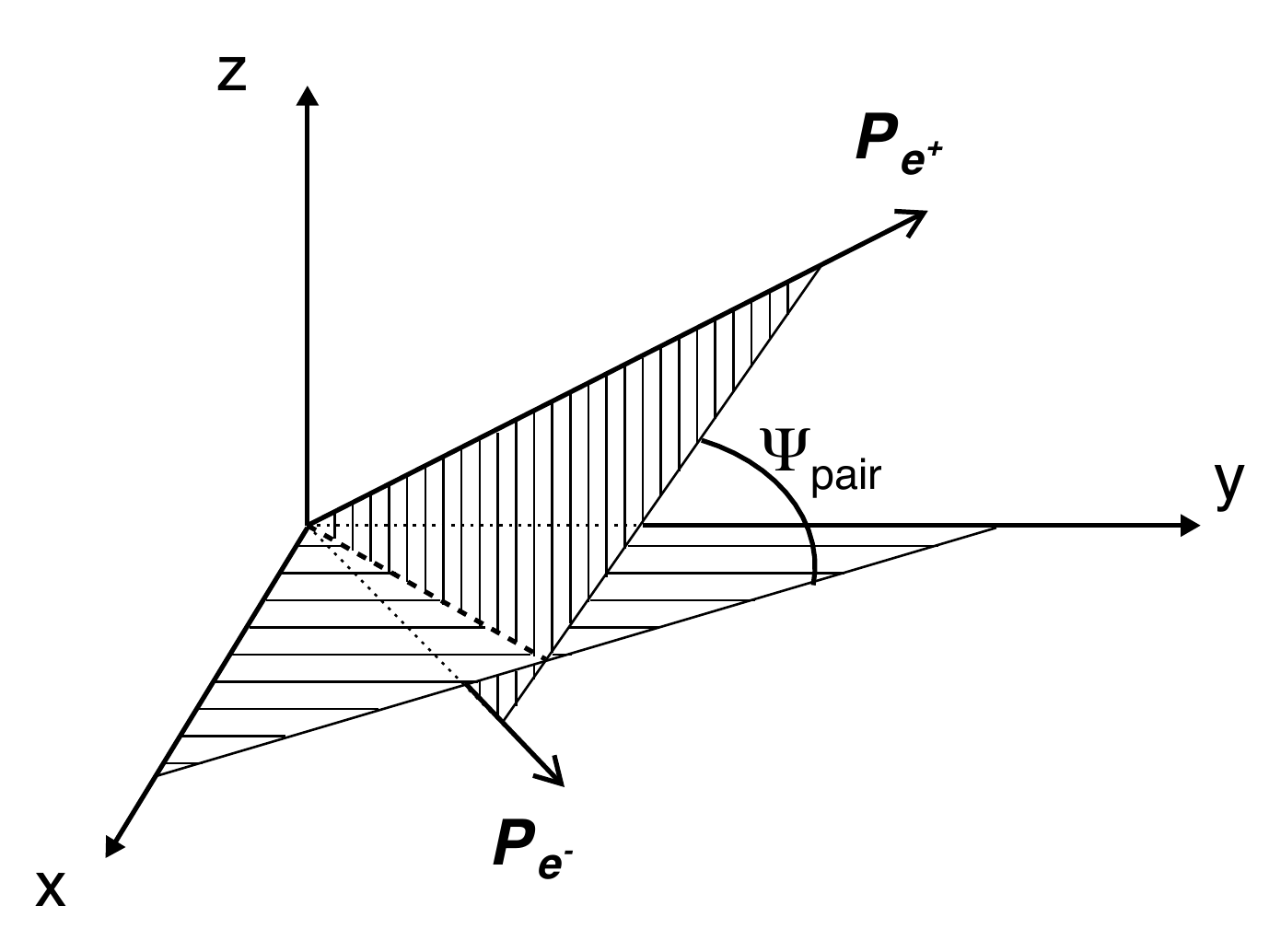}\label{fig_gamConvB}}
			\put(-30,10){(b)}
			\caption{(a): Projections of $e^+$ and $e^-$ tracks in the $x$-$y$ plane. The $z$-axis shows the direction of the magnetic field. The reconstructed vertex of the tracks is near the IP (since the IP is used in the determination of the track helix parameters). Point CP is the conversion point obtained with the PCF, and is supposed to be the true vertex of the tracks. The distance from the IP to CP is $R_{xy}$. The points $O_1$, $O_2$ are the centers of the two track projections. The points A, B are the intersection points of $O_1O_2$ and the two track projections, and the distance between A and B is $|\Delta xy|$. The arrow \bm{$P$}$_{\gamma}$ represents the momentum of the converted photon, and the angle between the arrow and the IP-CP is $\theta_{eg}$. (b): Illustration of $\Psi_{\rm{pair}}$. The arrow \bm{$P$}$_{e^+}$ (\bm{$P$}$_{e^-}$) represents the momentum of the $e^+$ ($e^-$).}
		\end{center}
		\end{figure}

		The radiative photon is reconstructed from the $e^+e^-$ pair using the PCF. At BESIII, the helix parameters of charged tracks are determined assuming that the IP is the origin, which is not true in our case since the conversion point (CP) is generally displaced from the IP. The conversion point of the photon is estimated using the track projections of $e^+e^-$ in the $x$-$y$ plane, perpendicular to the beam direction. The midpoint of the centers of the two track projections is taken as the CP, as shown in Fig.~\ref{fig_gamConvA}. As most photon conversions occur at the beam pipe and the inner wall of the MDC, the distances from the CP to IP in the $x$-$y$ plane, denoted by $R_{xy}$, is usually greater than 2~cm. Hence, $R_{xy}>2$~cm is required to suppress non-conversion $e^+e^-$ pairs. Moreover, as the radiative photon has high energy, the opening angle between the conversion $e^+$ and $e^-$ tracks is close to zero. Based on this, several selection criteria are applied to suppress the non-conversion $e^+e^-$ tracks: (i) The sum of the minimum distances from the CP to the two track projections, denoted by $|\Delta xy|$, has to be less than 0.2~cm, see Fig.~\ref{fig_gamConvA}. (ii) The minimum distance between the tracks of $e^+e^-$ in the beam direction, denoted by $\Delta z$, has to be less than 1.5~cm. (iii) The angle between the $x$-$y$ plane and the plane determined by the momentum vectors of $e^+$ and $e^-$, denoted by $\Psi_\text{pair}$, has to be within $[-0.5, 0.5]$ radians, see Fig.~\ref{fig_gamConvB}. The $\Psi_\text{pair}$ of converted $e^+e^-$ concentrate around zero because their polar angles are essentially the same, but their azimuth angles are slightly different as the track parameters are extrapolated to the IP rather than the CP. (iv) The angle between the momentum vector of the radiative photon and the direction from IP to CP, denoted by $\theta_{eg}$, has to satisfy $\!\cos\theta_{eg}>0.8$, see Fig.~\ref{fig_gamConvA}.

		After the reconstruction of the radiative photon, further selection criteria are applied to suppress background events. Since almost all known $\eta$ decays contain at least one photon, we require that at least one photon is detected in the EMC to suppress fully charged background contributions. The photon candidate must have a deposited energy greater than 25 MeV when detected in the barrel region ($|\!\cos\theta|<0.80$) and greater than 50 MeV when detected in the end cap region ($0.86<|\!\cos\theta|<0.92$). The angle between the detected position of the photon candidate and the closest extrapolated charged track must be larger than 10 degrees to exclude photons that originate from charged tracks. The difference between the EMC time of the photon candidate and the event start time is required to be within [0, 700] ns to suppress electronic noise and photons unrelated to the event. Furthermore, to suppress $e^+e^-\to\gamma\gamma$ and $J/\psi\to\gamma_1X,~X\to\gamma_2\gamma_3$, where $\gamma_2$ or $\gamma_3$ converts to $e^+e^-$, three selection criteria are applied. (i) The energy of all photons (except for the radiative one) are required to be less than 1.4 GeV. (ii) For events that have fewer than five photons, $-0.998<\!\cos\theta_{\gamma\gamma}<0$ is required, where $\theta_{\gamma\gamma}$ is the angle between the radiative photon and the most energetic one of the other photons. (iii) For events that have only two charged tracks and fewer than four photons, $|\!\cos\theta_{\rm{miss}}|<0.98$ is required, where $\theta_{\rm{miss}}$ is the polar angle of the missing momentum of the event. Finally, to suppress $J/\psi\to e^+e^-(\gamma),~\gamma\to e^+e^-$ events, all events that have more than two charged tracks are required to satisfy $2P_{\rm{trk}}-P_{\gamma}<0.8$~GeV, where $P_{\gamma}$ and $P_{\rm{trk}}$ are the magnitude of the momentum of the radiative photon and the most energetic charged track excluding the converted $e^+e^-$, respectively.

		According to a study with the MC sample, only 0.06\% of the $J/\psi\to\gamma\eta$ events that passed all of the above selection criteria have more than one $e^+e^-$ combination. All combinations are retained for further analysis.

		\begin{figure}[htbp]
		\begin{center}
			\includegraphics[width=\columnwidth]{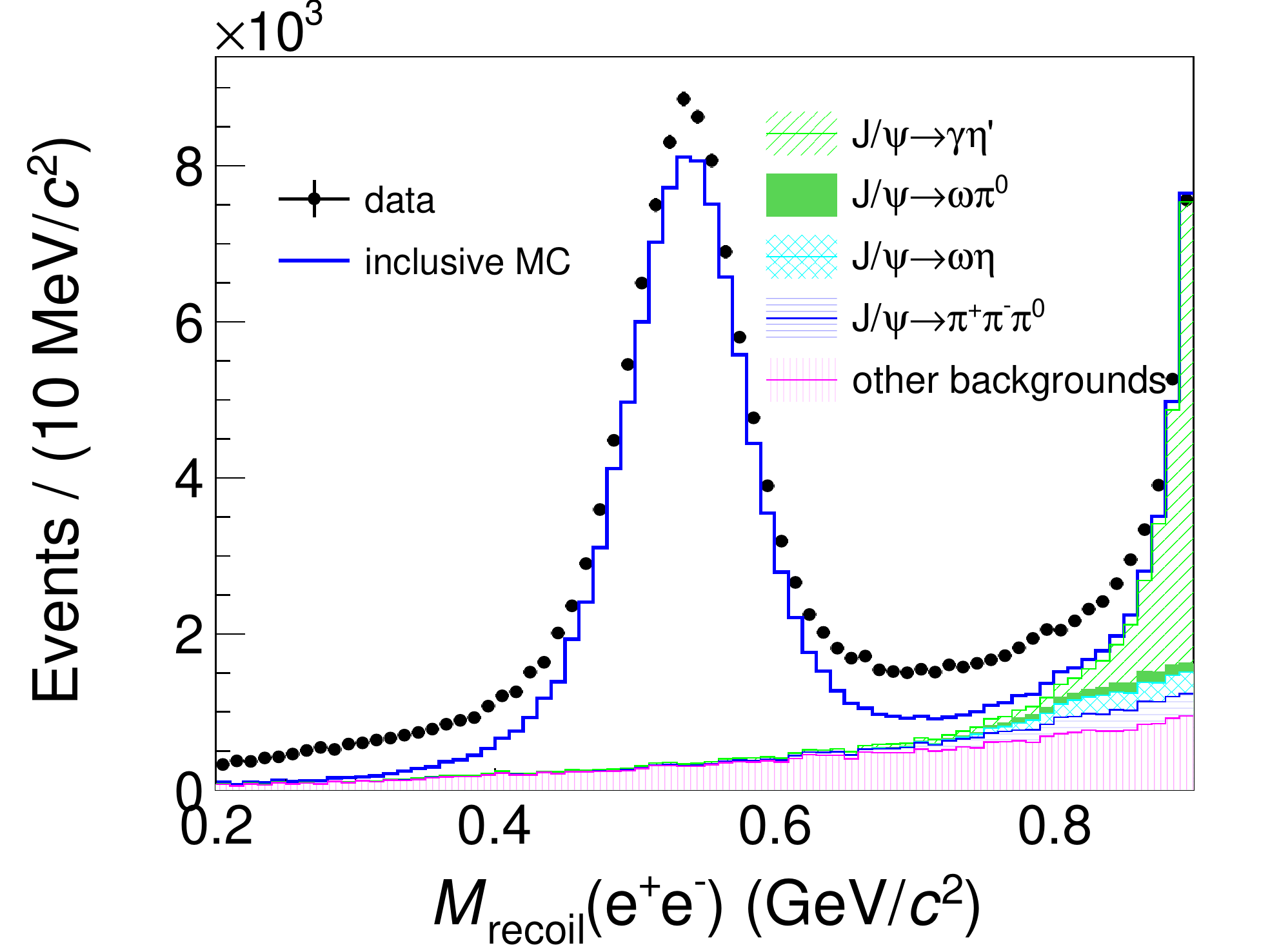}
			\caption{Recoil mass spectrum of $e^+e^-$ for the final event sample. The black dots with error bars represent data and the solid blue line represents the inclusive MC sample. The five shaded (or solid) histograms represent five different background components from the inclusive MC sample and are drawn stacked upon each other. The histogram filled with light green oblique lines represents the $J/\psi\to\gamma\eta'$ events, the green filled solid histogram $J/\psi\to\omega\pi^0$ events, the histogram filled with light blue grids $J/\psi\to\omega\eta$ events, the histogram filled with dark blue horizontal lines $J/\psi\to\pi^+\pi^-\pi^0$ events, and the histogram filled with pink vertical lines the other background events.}\label{fig_mEtaDataInc}
		\end{center}
		\end{figure}

		The recoil mass spectra of $e^+e^-$ for data and the corresponding inclusive MC sample, after all the selection criteria have been applied, are shown in Fig.~\ref{fig_mEtaDataInc}. There are large differences between the two samples. The reason is that some background processes are not included in the inclusive MC sample. These missing backgrounds are from the processes $e^+e^-\to\gamma\gamma$, $e^+e^-\to e^+e^-$ and $J/\psi\to e^+e^-\eta$. These processes are instead studied with exclusive MC samples. In addition, to describe the background events more accurately, some processes already included in the inclusive MC sample are simulated exclusively. These processes are $J/\psi\to\gamma\eta'$, $J/\psi\to\pi^+\pi^-\pi^0$, $J/\psi\to\omega\eta$, and $J/\psi\to\omega\pi^0$. The remaining backgrounds are studied with the inclusive MC sample. Different background components from the inclusive MC sample are shown in ~\ref{fig_mEtaDataInc}. The background caused by wrong $e^+e^-$ combinations of $J/\psi\to\gamma\eta$ events is ignored.

	\subsection{\label{sec:exclusive_selections}Exclusive channels}
		Candidate events for the processes $J/\psi\!\to\!\gamma\eta$, $\eta\!\to\!X$ ($X\!=\!\gamma\gamma$, $\pi^0\pi^0\pi^0$, $\pi^+\pi^-\pi^0$ or $\pi^+\pi^-\gamma$) are reconstructed with the following common selection criteria. (i) Charged tracks detected in the MDC are required to have a polar angle $|\!\cos\theta|<0.93$, and the distance of closest approach to the IP must be less than 10~cm along the beam direction and less than 1~cm in the transverse plane. (ii) Photons are reconstructed with the same selection criteria as described in section~\ref{sec:inclusive_selections}, except that only photons detected in the barrel region ($|\!\cos\theta|<0.80$) of the EMC are used, where the photon detection efficiency of data is in good agreement with that of MC simulation. In addition, for the neutral decays of $\eta\to\gamma\gamma$ and $\eta\to\pi^0\pi^0\pi^0$, instead of being within [0, 700] ns of the event start time, the EMC times of the photons are required to be within $[-500, 500]$~ns of the EMC time of the most energetic photon. (iii) The events must have the correct number of charged tracks, and at least the minimum number of photons associated with the given final state. (iv) A kinematic fit on the final state particle candidates is performed. The kinematic fit adjusts the track energy and momentum within the measured uncertainties so as to satisfy energy and momentum conservation for the given final state hypothesis. This improves the momentum resolution and reduces the background. (v) To maximize the figure of merit, defined as $S/\sqrt{S+B}$, the maximum value of the kinematic fit quality of the candidate events, $\chi^2$, is restricted. Here, $S$ is the number of corresponding signal events estimated by MC simulation, and $S+B$ is the number of data events. (vi) If there are multiple possible photon combinations, the combination with the minimum $\chi^2$ is chosen for further analysis. (vii) For the channels $X=\pi^0\pi^0\pi^0$, $\pi^+\pi^-\pi^0$, and $\pi^+\pi^-\gamma$, the energy of the radiative photon is much larger than that of the other photons. Therefore, the most energetic photon is taken as the radiative photon.

\begin{figure*}[htbp]
		\begin{center}
			\subfigure{\includegraphics[width=\columnwidth]{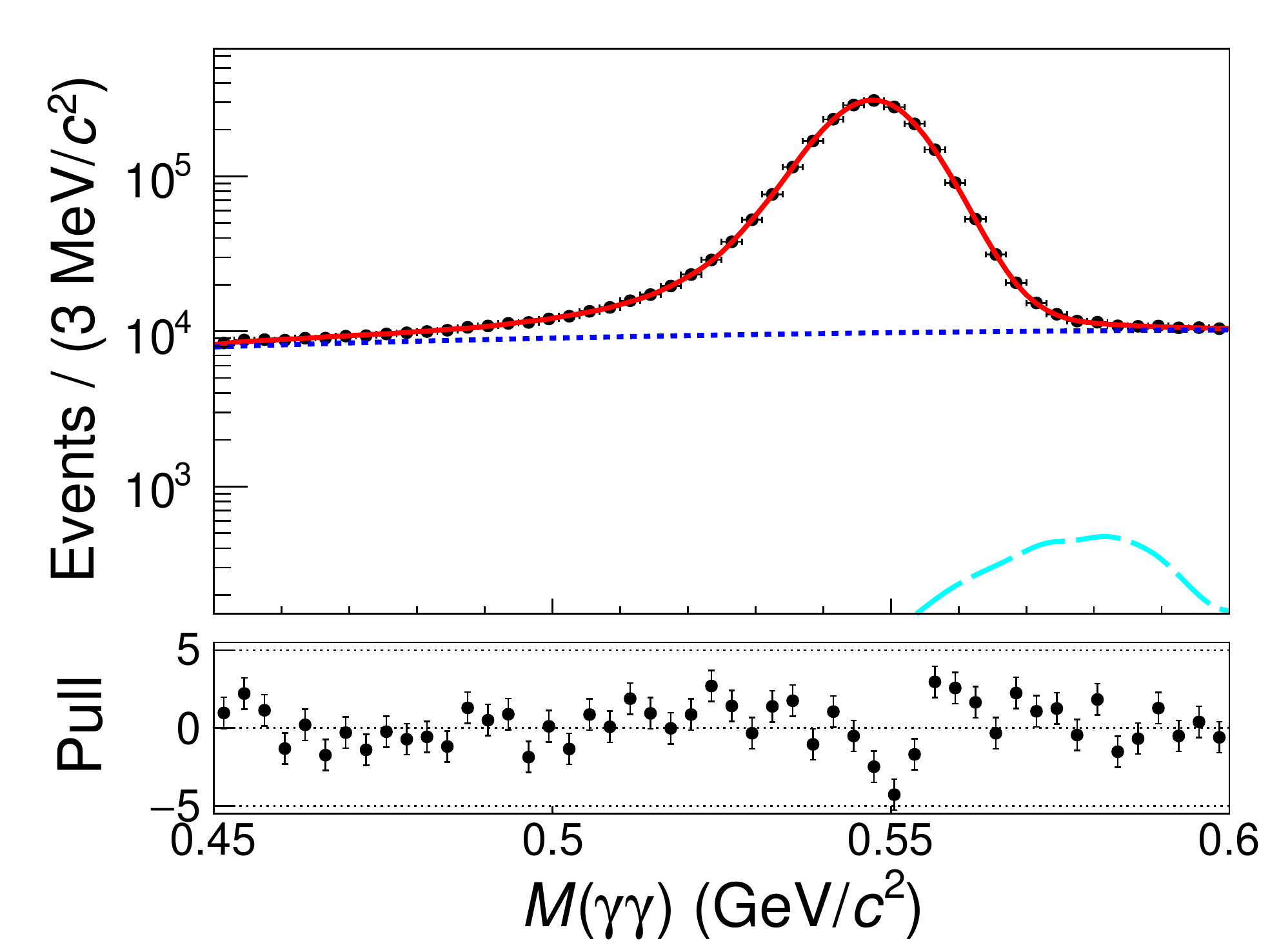}\label{fig_fitC0}}
			\put(-40,150){(a)}
			\subfigure{\includegraphics[width=\columnwidth]{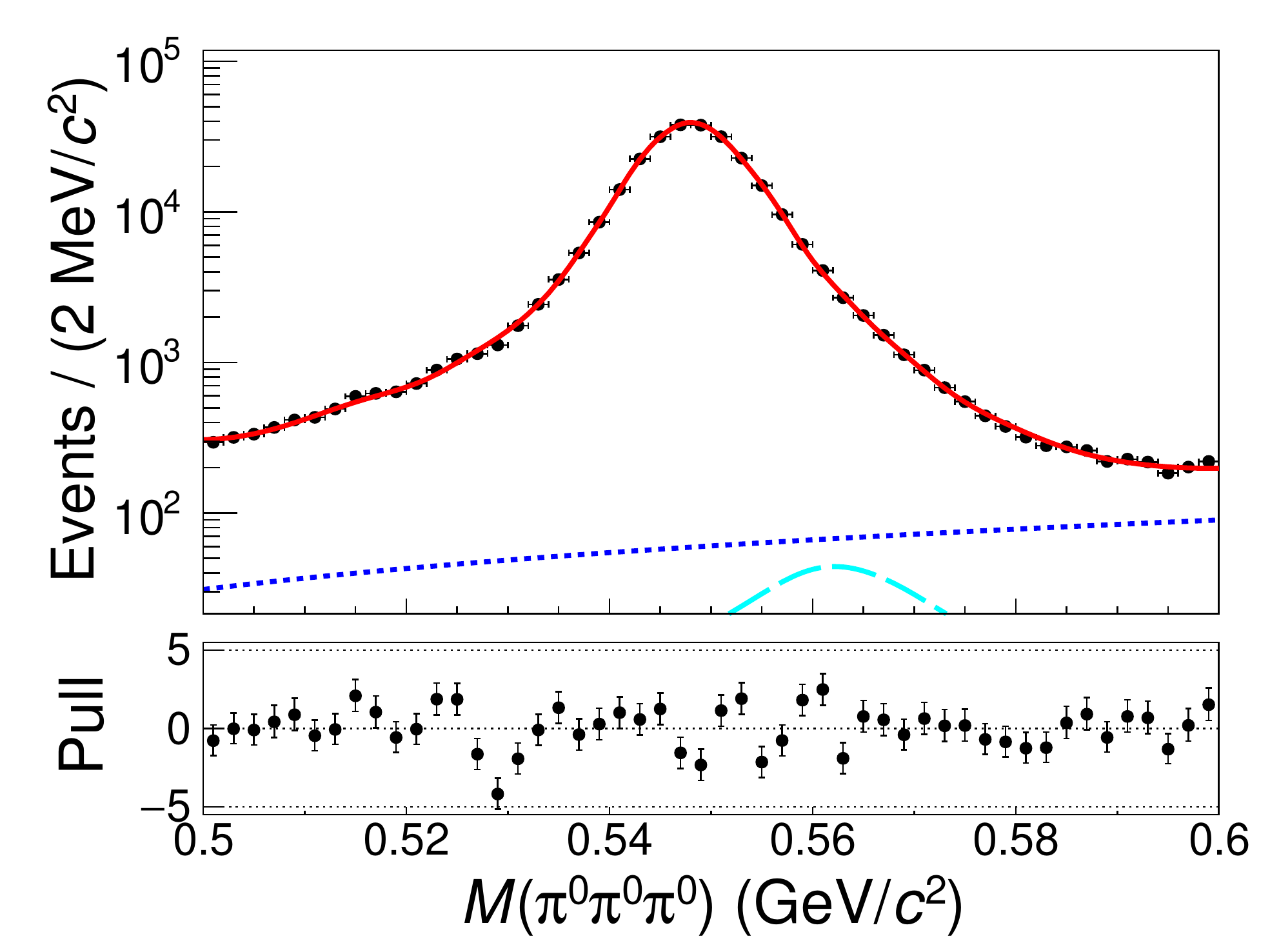}\label{fig_fitC1}}
			\put(-40,150){(b)}\\
			\subfigure{\includegraphics[width=\columnwidth]{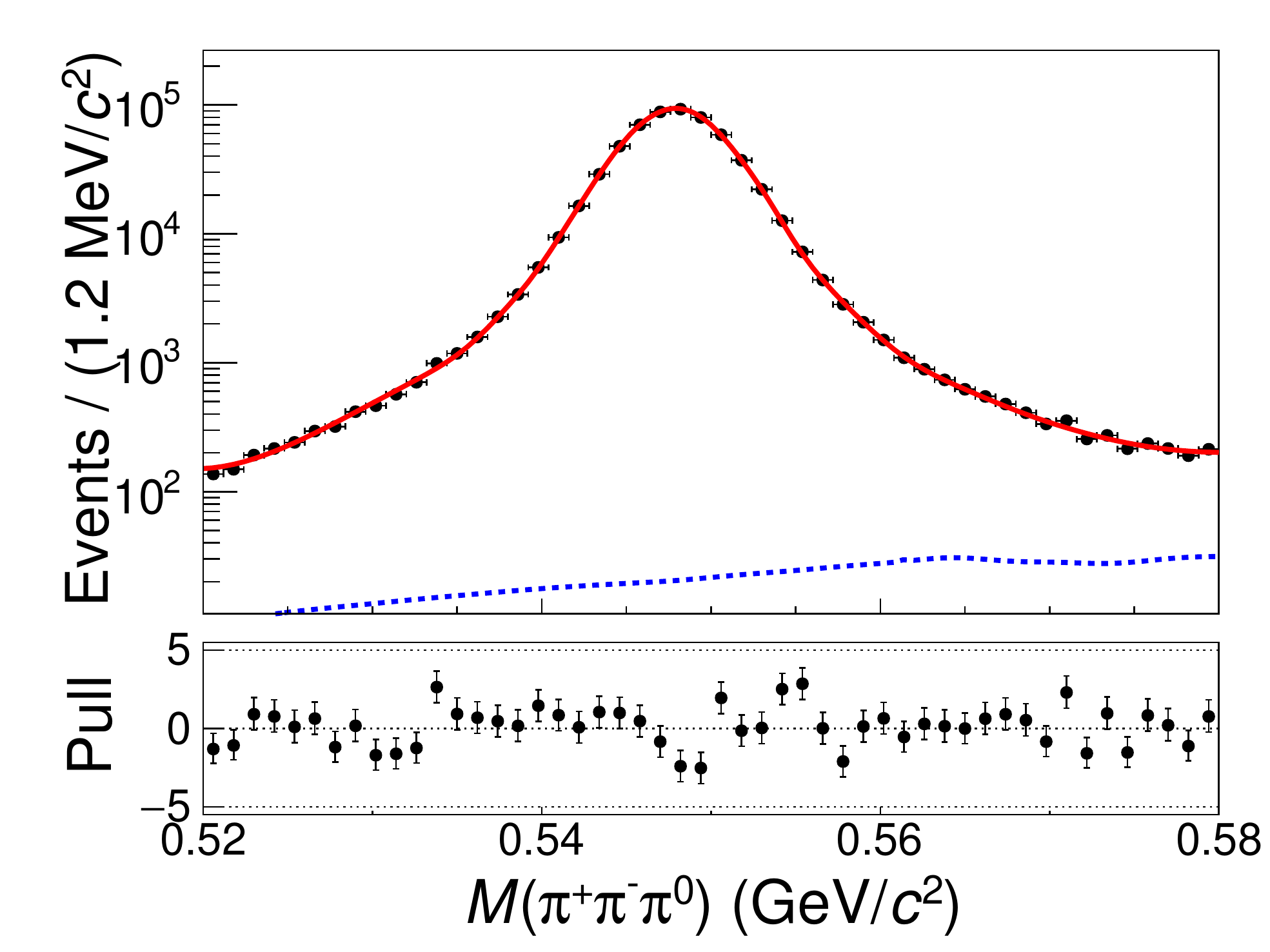}\label{fig_fitC3}}
			\put(-40,150){(c)}
			\subfigure{\includegraphics[width=\columnwidth]{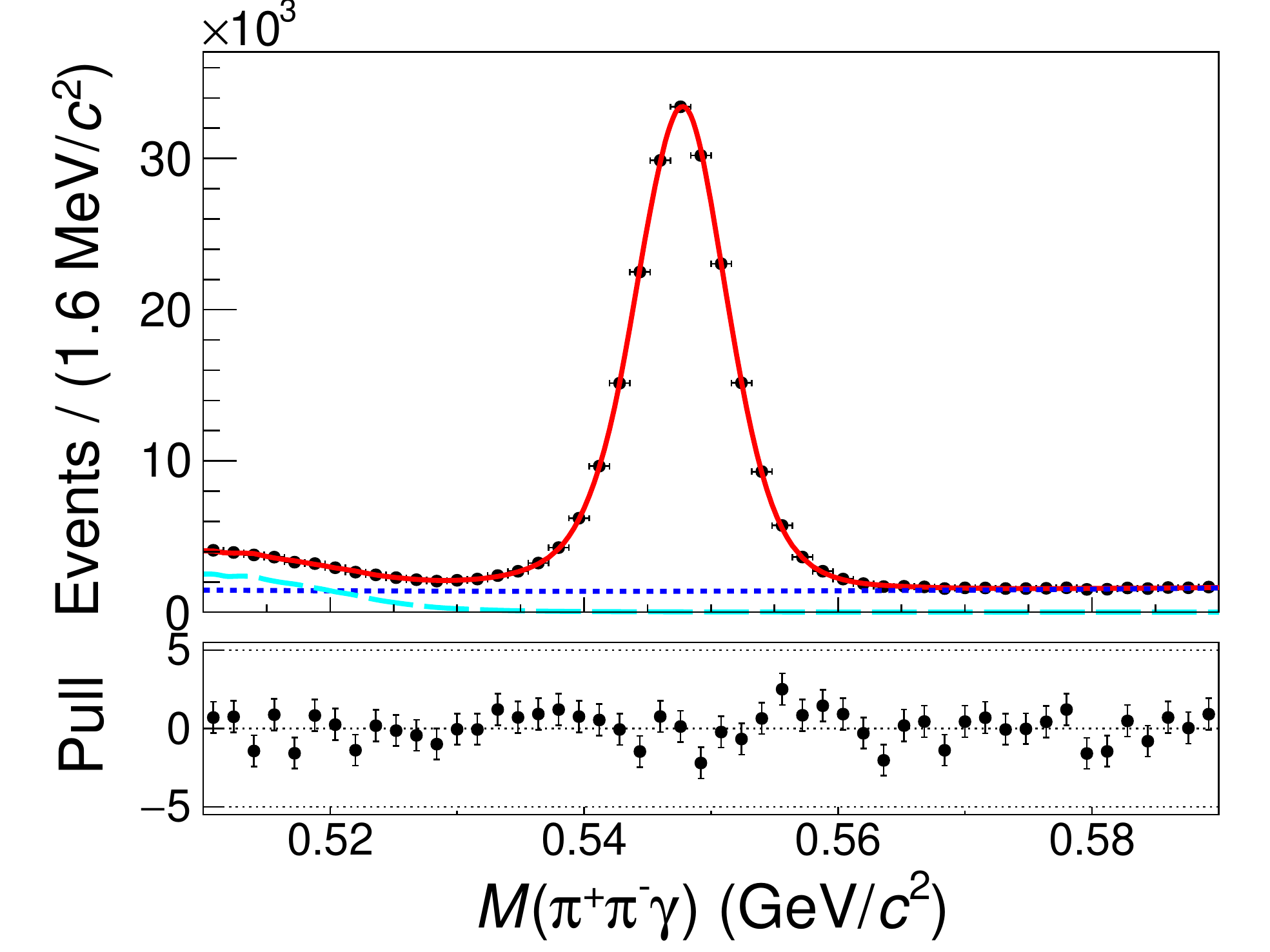}\label{fig_fitC2}}
			\put(-40,150){(d)}
			\caption{Unbinned maximum likelihood fit to the mass spectra of the $\eta$ decay modes. The black dots with error bars represent data, the red solid lines represent the fit results, the light blue dashed lines represent the peaking backgrounds, and the dark blue dotted lines represent the other backgrounds. The bottom panel shows the corresponding pull distribution. (a) The $\eta\to\gamma\gamma$ channel. (b) The $\eta\to\pi^0\pi^0\pi^0$ channel. (c) The $\eta\to\pi^+\pi^-\pi^0$ channel. (d) The $\eta\to\pi^+\pi^-\gamma$ channel.}\label{fig_fitC}
		\end{center}
		\end{figure*}

		In the case of $J/\psi\to\gamma\eta,~\eta\to\gamma\gamma$, a four-constraint (4C) kinematic fit imposing energy-momentum conservation is performed, and the fit quality $\chi^2_{\rm{4C}}$ is required to be less than 80. To suppress the $e^+e^-\to\gamma\gamma(\gamma)$ process, the energy of the photons is required to be greater than 0.07 GeV. The $\eta$ is reconstructed using $\gamma\gamma$ pairs, shown in Fig.~\ref{fig_fitC0}. As it is impossible to separate the radiative photons from the $\eta$-decay photons, all $\gamma\gamma$ combinations are kept. MC simulations of the signal show that the mass spectrum of the wrong $\gamma\gamma$ combinations is flat. 
		In addition, the background distributions have been investigated with the inclusive MC sample. Except for events from the processes $J/\psi\to\omega\eta,~\omega\to\gamma\pi^0,~\eta\to\gamma\gamma$, and $J/\psi\to\gamma f_0(2100),~f_0(2100)\to\eta\eta,~\eta\to\gamma\gamma$, which form a small peak in the signal region, the distribution of the other background contributions is smooth.

		For the decay $J/\psi\to\gamma\eta$,~$\eta\to\pi^0\pi^0\pi^0$, a seven-constraint (7C) kinematic fit imposing energy-momentum conservation and constraints on the three $\pi^0$ masses is performed. The fit $\chi^2_{\rm{7C}}$ is required to be less than 100. The three-$\pi^0$ combination with the least $\chi^2_{\rm{7C}}$ is used to reconstruct the $\eta$, as displayed in Fig.~\ref{fig_fitC1}. A very clean $\eta$ peak is observed. 
		Using the inclusive MC sample of $J/\psi$ decays, the background study indicates that only the decays $J/\psi\to\omega\eta,~\omega\to\gamma\pi^0,~\eta\to3\pi^0$, and $J/\psi\to\gamma f_0(2100),~f_0(2100)\to\eta\eta,~\eta\to3\pi^0$ may contribute to a very small peak in the signal region, as indicated by the dashed line in Fig.~\ref{fig_fitC1}.

		The $J/\psi\to\gamma\eta$, $\eta\to\pi^+\pi^-\pi^0$ candidates are selected with a five-constraint (5C) kinematic fit imposing energy-momentum conservation and a constraint on the mass of the $\pi^0$, and the $\chi^2_{\rm{5C}}$ is required to be less than 100. 
		After the above requirements, the $\pi^+\pi^-\pi^0$ invariant mass is illustrated in Fig.~\ref{fig_fitC3}, where a prominent $\eta$ peak is seen. We also perform a background study with the inclusive MC sample, and the result indicates that no peaking background is seen in the $\eta$ mass region.

		The $J/\psi\to\gamma\eta$, $\eta\to\pi^+\pi^-\gamma$ candidates are selected using a 4C kinematic fit, and the $\chi^2_{\rm{4C}}$ is required to be less than 60. If more than two good photons are found, a 5C kinematic fit under the $J/\psi\to\gamma\eta$, $\eta\to\pi^+\pi^-\pi^0$ hypothesis is performed. After requiring that the kinematic fit probability of $\gamma\pi^+\pi^-\gamma$ is greater than that under the hypothesis of $\gamma\pi^+\pi^-\pi^0$, the mass spectrum of $\pi^+\pi^-\gamma$ is shown in Fig.~\ref{fig_fitC2}. The MC simulation shows that the background events from $\eta\to\pi^+\pi^-\pi^0$ may contribute to a broad bump on the left side of the $\eta$ peak, while no peaking background events are found in the signal region.

                \section{Measurement of $\boldmath\mathcal{B}(J/\psi\to\gamma\eta)$}
		To measure the BF of the $J/\psi\to\gamma\eta$ decay, an unbinned maximum likelihood fit is performed to the recoil mass spectrum of the $e^+e^-$ pair. The fit includes a signal component and background contributions estimated from exclusive and inclusive simulations.

		The distribution of the signal events is described by a modified double-tailed Crystal Ball function:
		\begin{equation}\label{eq:modifiedCB}
			X\!=\!\frac{x\!-\!\mu}{\sigma},~f(x)\!=\!
			\begin{cases}
				\frac{\rm{exp}\left(-\frac{1}{2}|k|^{1+\frac{1}{1+0.5|k|}}\right)}{(1-\frac{1}{n}-\frac{|k|}{n}X)^{n}}, & X\!<\!-|k|\\\\
				\rm{exp}\left(-\frac{1}{2}|X|^{1+\frac{1}{1+0.5|X|}}\right), & \!-|k|\!\leq\!  X\!\leq\! |K|\\\\
				\frac{\left(-\frac{1}{2}|K|^{1+\frac{1}{1+0.5|K|}}\right)}{(1-\frac{1}{N}+\frac{|K|}{N}X)^{N}}, & X>|K|
			\end{cases}
		\end{equation}
		Compared to the standard Crystal Ball function~\cite{CBShape}, it uses a modified Gaussian function as the core region. The parameters $n, k, N$, and $K$ describe the two tails, while $\mu$ and $\sigma$ are parameters of the modified Gaussian function. We first fit the signal MC sample with Eq.~(\ref{eq:modifiedCB}), where all the parameters are included in the fit. In the second step, a fit to data is performed, in which the values of $n, k, N$, and $K$ are fixed to the results obtained by fitting the signal MC sample. The parameters $\mu$ and $\sigma$ are determined by the fit to data.

		Background events from the processes $e^+e^-\to\gamma\gamma$, $e^+e^-\to e^+e^-$, $J/\psi\to e^+e^-\eta$, $J/\psi\to\pi^+\pi^-\pi^0$, $J/\psi\to\omega\eta$, $J/\psi\to\omega\pi^0$, and $J/\psi\to\gamma\eta'$ are described with shapes extracted from their corresponding exclusive MC samples. The number of $J/\psi\to\gamma\eta'$ events is left free in the fit, while the numbers of events from the other six processes are fixed according to $N=L\sigma\varepsilon$ or $N=N_{J/\psi}\mathcal{B}\varepsilon$. Here, $L$ is the integrated luminosity of the data sample, $N_{J/\psi}$ is the number of $J/\psi$ events, $\varepsilon$ stands for the efficiency estimated with the corresponding MC sample, and $\sigma$ (or $\mathcal{B}$) stands for the corresponding cross section (or BF). The BFs of the $J/\psi\to e^+e^-\eta$ and $J/\psi\to\pi^+\pi^-\pi^0$ decays are obtained from Ref.~\cite{eeEtaBF} and~\cite{rhoPiBF2}. The BFs of $J/\psi\to\omega\eta$ and $J/\psi\to\omega\pi^0$ are obtained from Ref.~\cite{omegaEtaBF}.

\begin{figure}[htbp]
		\begin{center}
			\includegraphics[width=\columnwidth]{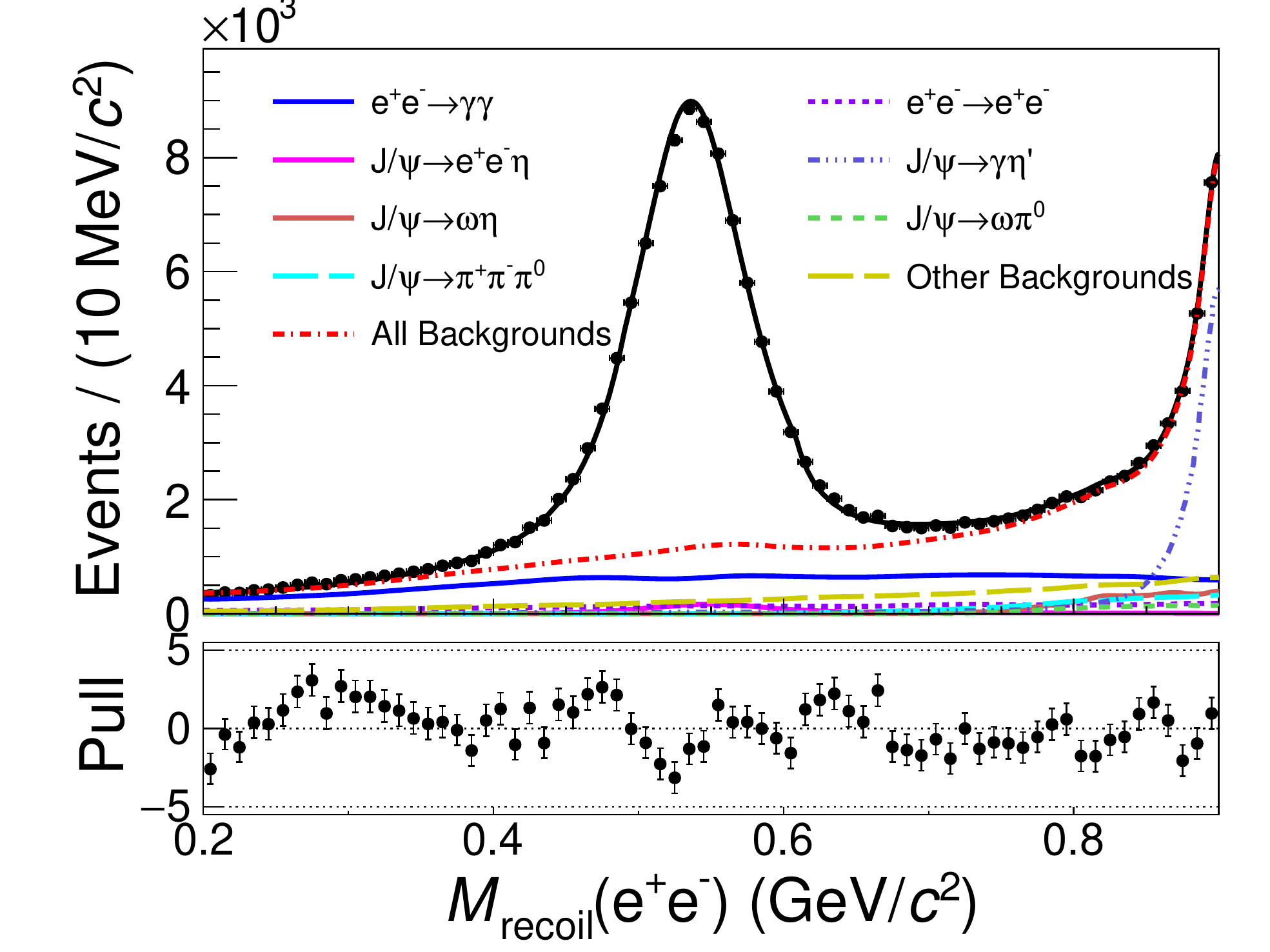}
			\caption{Unbinned maximum likelihood fit to the $e^+e^-$ recoil mass spectrum. In the top panel, the black dots with error bars are data. The black solid curve gives the fit result. The red dashed line represents the distribution of all backgrounds. The blue solid, purple dotted, pink solid, dark blue dashed, brown solid, green dotted, light blue dashed, and yellow dashed lines represent the backgrounds from $e^+e^-\to\gamma\gamma$, $e^+e^-\to e^+e^-$, $J/\psi\to e^+e^-\eta$, $J/\psi\to\gamma\eta'$, $J/\psi\to\omega\eta$, $J/\psi\to\omega\pi^{0}$, $J/\psi\to\pi^+\pi^-\pi^0$, and the other backgrounds, respectively. The bottom panel shows the corresponding pull distribution.}\label{fig_fitInc}
		\end{center}
		\end{figure}

		The remaining background events are described using the shape extracted from the inclusive MC sample. The normalization of this component is determined by the fit. The result of the fit is shown in Fig.~\ref{fig_fitInc}.

		In the last step, a correction factor $f$, which accounts for the difference in the photon conversion efficiency between data and MC simulation, is implemented in the BF calculation. Using $e^+e^-\to\gamma\gamma$ events collected at $\sqrt{s}=3.08$ GeV, just below the $J/\psi$ resonance, the factor $f=\Fconv\pm\FconvEsta\pm\FconvEsys$ is estimated with
		\begin{equation}\label{eq:fConv}
			f\equiv\frac{ N^{\rm{data}}_{\rm{conv}}/N^{\rm{data}}_{\gamma\gamma} }{ N^{\rm{mc}}_{\rm{conv}}/N^{\rm{mc}}_{\gamma\gamma} }\approx\frac{ \varepsilon^{\rm{data}}_{\rm{conv}} }{ \varepsilon^{\rm{mc}}_{\rm{conv}} },
		\end{equation}
where $N_{\rm{conv}}$ and $N_{\gamma\gamma}$ are the observed numbers of $e^+e^-\to\gamma\gamma$ events with and without one $\gamma\to e^+e^-$ conversion, respectively. The energy of the radiative photons in the $e^+e^-\to\gamma\gamma$ sample and the $J/\psi\to\gamma\eta$ sample are very close.

		Both the reconstruction of the converted and non-converted photons are the same as described above. The systematic uncertainty of $f$ is conservatively estimated to be $(f-1)/2$. The yields of $J/\psi\to\gamma\eta$, $\gamma\to e^+e^-$ events is $N_{\gamma\eta}^{\rm{obs}}=87887\pm373$. The detection efficiency obtained from MC simulation is $\varepsilon_{\gamma\eta}=(\effInc\pm\effIncErr)\times 10^{-3}$, and $\mathcal{B}(J/\psi\to\gamma\eta)$ is determined to be $(\brInc\pm\brIncEsta)\!\times\!{10^{-3}}$, where the uncertainty is statistical.

	\section{Measurement of $\boldmath \mathcal{B}(\eta\to X)$}

		The signal yields of the exclusive channels are obtained by performing unbinned maximum likelihood fits to the mass spectra of $\gamma\gamma$, $\pi^0\pi^0\pi^0$, $\pi^+\pi^-\pi^0$, and $\pi^+\pi^-\gamma$ candidates, respectively. In the fits, the signal component is modeled by the MC-simulated shape convolved with a Gaussian function to account for the difference in the mass resolution between MC simulation and data. The parameters of the Gaussian function are free in the fit. When simulating the signal shape of the $\eta\to\gamma\gamma$ channel, only the right $\gamma\gamma$ combination of the MC events is used. The wrong $\gamma\gamma$ combinations are removed according to the simulation information.

		The backgrounds of the $\eta\to\gamma\gamma$ and $\eta\to\pi^0\pi^0\pi^0$ channels are modeled with two components: (i) a second-order Chebyshev polynomial function which describes the non-peaking background; (ii) a peaking background shape which is determined with MC simulation, and the number of the peaking background events is fixed according to the corresponding BF from PDG~\cite{pdg2019}. The background of the $\eta\to\pi^+\pi^-\pi^0$ channel is modeled by the shape obtained from the inclusive MC sample. The background of the $\eta\to\pi^+\pi^-\gamma$ channel is modeled by the sum of a second-order Chebyshev polynomial function and the shape obtained from the $J/\psi\to\gamma\eta$, $\eta\to\pi^+\pi^-\pi^0$ MC sample. Here the magnitudes of different components are left free in the fit.
		The fit results of the four channels are shown in Fig.~\ref{fig_fitC}.

		The signal yields obtained from the fits, the detection efficiencies estimated with MC simulations, and the BFs of the four dominant $\eta$ decay modes are listed in Table~\ref{table_sumBF}. Note that the BFs from CLEO and the PDG are all relative measurements.

		\begin{table*}[htbp]
		\begin{center}
			\caption{Summary of BFs and comparison with previous results. The first error is statistical and the second systematic.}\label{table_sumBF}
			\setlength{\tabcolsep}{0.2cm}{\begin{tabular}{lllrrr}\hline\hline
				& & & \multicolumn{3}{c}{$\mathcal{B}(\eta\to X)~(\%)$}\\\cline{4-6}
				\multicolumn{1}{c}{$X$} & \multicolumn{1}{c}{$N^{\rm{obs}}_{\eta\to X}$($\times 10^5$)} & \multicolumn{1}{c}{$\varepsilon_{\gamma\eta}$(\%)} & \multicolumn{1}{c}{This Work} & \multicolumn{1}{c}{CLEO} & \multicolumn{1}{c}{PDG}\\\hline
				$\gamma\gamma$ & 20.78$\pm$0.02 & $\effCa\pm\effCaErr$ & $\brCa\pm\brCaEsta\pm\brCaEsys$ & 38.45$\pm$0.40$\pm$0.36 & 39.41$\pm$0.20\\
				$\pi^0\pi^0\pi^0$ & 2.831$\pm$0.006 & $\effCb\pm\effCbErr$ & $\brCb\pm\brCbEsta\pm\brCbEsys$ & 34.03$\pm$0.56$\pm$0.49 & 32.68$\pm$0.23\\
				$\pi^+\pi^-\pi^0$ & 6.131$\pm$0.008 & $\effCd\pm\effCdErr$ & $\brCd\pm\brCdEsta\pm\brCdEsys$ & 22.60$\pm$0.35$\pm$0.29 & 22.92$\pm$0.28\\
				$\pi^+\pi^-\gamma$ & 2.018$\pm$0.005 & $\effCc\pm\effCcErr$ & $\brCc\pm\brCcEsta\pm\brCcEsys$ & 3.96$\pm$0.14$\pm$0.14 & 4.22$\pm$0.08\\
				\hline\hline
			\end{tabular}}
		\end{center}
		\end{table*}

	\section{Systematic Uncertainties}\label{sec:sysU}
		The systematic uncertainties of $\mathcal{B}(J/\psi\to\gamma\eta)$ have been evaluated for the fit procedure, the reconstruction efficiency of the converted photon, the efficiencies of the background suppression criteria, and the number of $J/\psi$ candidates. The systematic uncertainties of $\mathcal{B}(\eta\to X)$ have been evaluated for the fit procedure, the photon detection efficiency, the tracking efficiencies of charged pions, the kinematic fit efficiency, the efficiencies of the background suppression criteria, and the BFs of the decays $\pi^0\to\gamma\gamma$ and $J/\psi\to\gamma\eta$. As the number of $J/\psi$ canceles when calculating $\mathcal{B}(\eta\to X)$, the systematic uncertainty of $\mathcal{B}(\eta\to X)$ does not contain the uncertainty of the number of $J/\psi$. The reconstruction efficiencies cannot be cancelled as the radiative photons are reconstructed in different ways.

		The fit uncertainty comes from three sources: the fit range, the signal shape, and the background shape.
		The uncertainty arising from the fit range is estimated by varying the range. The change of the efficiency caused by the change of the fit range is considered. The change in the BF is taken as the systematic uncertainty.
		To estimate the uncertainty arising from the signal shape, we use the Bukin function~\cite{RooBukinLink2} instead of the MC-simulated shape to describe the signal component. The Bukin function is an asymmetric function with five parameters: $\mu,~\sigma,~\xi,~\rho_1$, and $\rho_2$. The parameters $\mu$ and $\sigma$ are the position and width of the peak, respectively, $\xi$ describes the asymmetry of the peak, and $\rho_1$, $\rho_2$ describe the tails of the peak. In the fit to the exclusive channels, all the parameters of the Bukin function are free. However, in the fit to the inclusive channel, we first fit the signal MC sample with all the parameters of the Bukin function free. Then we fit the data with values of $\rho_1$, $\rho_2$ fixed to the results of the fit to the signal MC sample, and $\mu,~\sigma,~\xi$ are free. The differences between the nominal results and the results from the alternative method are taken as the systematic uncertainties.

		The systematic effect arising from the background shapes is estimated with different methods for different channels. For the inclusive channel, two uncertainty sources are considered: the peaking background and the other background shape uncertainty. For the peaking background, the number of events is fixed during the fit. We vary the number by one standard deviation and the difference with respect to the nominal result is taken as the uncertainty. For the uncertainty of the remaining background, a 2nd-order Chebyshev function is added to the fit, and the induced change of the BF is taken as the uncertainty. The square root of the summed squares of the two uncertainties is taken as the background uncertainty.
		For the $\eta\to\pi^+\pi^-\pi^0$ channel, we use a 2nd-order Chebyshev function instead of the MC-simulated shape in the fit and take the change of the BF as the uncertainty. For the $\eta\to\gamma\gamma$, $\eta\to\pi^0\pi^0\pi^0$, and $\eta\to\pi^+\pi^-\gamma$ channels, the order of the Chebyshev polynomial functions used in the fit is changed and the induced change of the BF is chosen as the uncertainty. For the $\eta\to\gamma\gamma$ and $\eta\to\pi^0\pi^0\pi^0$ channels, the number of the peaking background events, which is fixed during the fit, is varied by one standard deviation, and the induced change of the BF is taken as one source of the uncertainty. The uncertainties from different sources are added in quadrature.

		As the reconstruction efficiency of the converted photon is corrected with the factor $f$, the uncertainty of $f$ is taken as the associated systematic uncertainty.
	
		For photons directly detected by the EMC, the uncertainty of the detection efficiency has been studied using a control sample of $e^+e^-\to\gamma\mu^+\mu^-$ events. The four-momentum of the initial-state-radiation photon is predicted using only the four-momentum of the $\mu^+\mu^-$ pair. The photon detection efficiency is defined as the fraction of predicted photons with four-momentum matching that of the actual photons in the EMC. The systematic uncertainty is defined as the relative difference in efficiency between data and MC simulation. It is found that the photon detection efficiency of data is consistent with the MC simulation within 0.5\%. The effect of the discrepancy between data and the MC simulation is estimated by using a reweighing technique. The weighted relative uncertainties for the $\eta\to\gamma\gamma$, $\eta\to\pi^0\pi^0\pi^0$, $\eta\to\pi^+\pi^-\pi^0$, and $\eta\to\pi^+\pi^-\gamma$ channels are determined to be $\sysGamOneCa$\%, $\sysGamOneCb$\%, $\sysGamOneCd$\%, and $\sysGamOneCc$\% per photon, respectively.

		The tracking efficiency uncertainty of charged pions has been studied with the control sample $J/\psi\to\pi^+\pi^-\pi^0$. The momentum of the $\pi^+$ is predicted using the four-momentum of $\pi^-\pi^0$, and the tracking efficiency of the $\pi^+$ is defined as the fraction of the number of events in which $\pi^+\pi^-\pi^0$ are reconstructed and the number of events in which $\pi^-\pi^0$ are reconstructed. The systematic uncertainty is defined as the relative difference in efficiency between data and MC simulation. The weighted average uncertainties for the tracks are obtained using bins of transverse momentum.
The weighted average relative uncertainties for the $\eta\to\pi^+\pi^-\pi^0$ and $\eta\to\pi^+\pi^-\gamma$ channels are \sysTrkOneCd\% and \sysTrkOneCc\% per track, respectively.

		The uncertainty associated with the kinematic fit arises from the inconsistency of the $\chi^2$ distribution between data and the MC simulation. The reconstructed energy and angle of the photons, the helix parameters of the charged tracks, and their errors of the MC simulation are corrected to make their distributions more consistent with data. This makes the $\chi^2$ distributions of data and the MC simulation more consistent as well. The corrected MC simulation is used for the nominal results. The difference of the kinematic fit efficiencies before and after the correction is taken as the uncertainty. The relative uncertainties for the four exclusive channels are $\sysKmfitCa$\%, $\sysKmfitCb$\%, $\sysKmfitCd$\%, and $\sysKmfitCc$\%, respectively.

		There are also efficiency uncertainties caused by the selection criteria used to suppress the background. For selection criteria that cause only a small efficiency loss, the corresponding uncertainties are conservatively estimated as half of the efficiency loss. Such selection criteria include $\chi^2_{\rm{4C}}<\chi^2_{\rm{5C}}$ for the $\eta\to\pi^+\pi^-\gamma$ channel, and the photon energy requirement $E_{\gamma}>0.07$ GeV for the $\eta\to\gamma\gamma$ channel.
		For the selection that requires at least one photon detected in addition to the radiative photon, which is applied on the inclusive channel, the change of the BF obtained with or without the selection is taken as the corresponding uncertainty.
		For other selection criteria, we vary the value of the selection criteria, and take the change of the BF as the corresponding uncertainty. Such selection criteria including $E_{\gamma}<1.4$ GeV, $-0.998<\!\cos\theta_{\gamma\gamma}<0$, $|\!\cos\theta_{\rm{miss}}|<0.98$, and $2P_{\rm{trk}}-P_{\gamma_{ee}}<0.8$ GeV, which are all applied on the inclusive channel.

		The uncertainty of the number of $J/\psi$ events is described in Sec.~\ref{sec_dataset}. Finally, the BF uncertainty of the $\pi^0\to\gamma\gamma$ decay is taken from the PDG~\cite{pdg2019}.

		The systematic uncertainties are summarized in Table~\ref{table_sysSumInc} and Table~\ref{table_sysSumEx}. In Table~\ref{table_sysSumInc}, the uncertainty of the number of $J/\psi$ events only contributed to the total uncertainty in the measurement of $\mathcal{B}(J/\psi\to\gamma\eta)$, as it can be canceled when calculating $\mathcal{B}(\eta\to X)$. The total systematic uncertainty is given by the quadratic sum of the individual contributions.

		\begin{table}[htbp]
		\centering
		\caption{Relative systematic uncertainties for the inclusive channel.}\label{table_sysSumInc}
		\begin{tabular}{lc}\hline\hline
			Source & Relative Uncertainty (\%)\\\hline
			Fit range & $\sysFitRInc$\\
			Signal shape & 0.57\\
			Background shape & $\sysBkgInc$\\
			Converted photon & $\FconvEPercent$\\
			At least one good photon & 0.61\\
			$E_{\gamma}<1.4$ GeV & 0.81\\
			$-0.998<\!\cos\theta_{\gamma\gamma}<0$ & 0.77\\
			$|\!\cos\theta_{\rm{miss}}|<0.98$ & 0.61\\
			$(2P_{\rm{trk}}-P_{\gamma})<0.8$ GeV & 0.41\\
			Number of $J/\psi$ events & 0.44\\\hline
			Total (for $\mathcal{B}(J/\psi\to\gamma\eta)$) & 2.24\\
			Total (for $\mathcal{B}(\eta\to X)$) & $\sysIncExSys$\\
			\hline\hline
		\end{tabular}
		\end{table}

		\begin{table}[htbp]
		\begin{center}
			\caption{Relative systematic uncertainties for the exclusive channels.}\label{table_sysSumEx}
			\begin{threeparttable}
			\begin{tabular}{lcccc}\hline\hline
				 & \multicolumn{4}{c}{Relative Uncertainty (\%)} \\\cline{2-5}
				Source\color{white}{\Large{1}} & $\gamma\gamma$ & $\pi^0\pi^0\pi^0$ & $\pi^+\pi^-\pi^0$ & $\pi^+\pi^-\gamma$\\\hline
				$\mathcal{B}(J/\psi\to\gamma\eta)$\tnote{*} & $\sysIncExSysSta$ & $\sysIncExSysSta$ & $\sysIncExSysSta$ & $\sysIncExSysSta$\\
				Fit range & $\sysFitRCa$ & $\sysFitRCb$ & $\sysFitRCd$ & $\sysFitRCc$\\
				Signal shape & 0.62 & 0.19 & 0.37 & 0.66\\
				Background shape & 0.39 & 0.28 & 0.04 & 0.13\\
				Photon detection efficiency & $\sysGamCa$ & $\sysGamCb$ & $\sysGamCd$ & $\sysGamCc$\\
				Tracking efficiency & -- & -- & $\sysTrkCd$ & $\sysTrkCc$\\
				Kinematic fit efficiency & $\sysKmfitCa$ & $\sysKmfitCb$ & $\sysKmfitCd$ & $\sysKmfitCc$\\
				Referenced BF & -- & {\color{white}0}0.059 & {\color{white}0}0.034 & --\\
				Other & 0.45 & -- & -- & 0.07\\\hline
				Total & $\reSysCa$ & $\reSysCb$ & $\reSysCd$ & $\reSysCc$\\
				\hline\hline
			\end{tabular}
			\begin{tablenotes}
			\footnotesize
			\item[*] Contains both the systematic and statistical uncertainties.
			\end{tablenotes}
			\end{threeparttable}
		\end{center}
		\end{table}

	\section{Summary}\label{sec:summary}
		Based on $(\nJpsi\pm\nJpsiErr)\times10^{10}$ $J/\psi$ events collected by BESIII at BEPCII, the BF of the decay $J/\psi\to\gamma\eta$ is measured with high precision and the absolute BFs of four dominant $\eta$ decays are measured for the first time. The measured BF of $J/\psi\to\gamma\eta$ is $(\brInc\pm\brIncEsta\pm\brIncEsys)\times 10^{-3}$, which is in agreement with the world average value, $(1.108\pm 0.027)\times 10^{-3}$~\cite{pdg2019}, within two standard deviations, but with improved precision.

		The measured BFs of $\eta$ decays are summarized in Table~\ref{table_sumBF}. The value of $\mathcal{B}(\eta\to\pi^+\pi^-\gamma)$ is consistent with the world average values~\cite{pdg2019} within two standard deviations, and the measured BFs of the other $\eta$ decays are within one standard deviation. Compared with the BFs measured by CLEO~\cite{cleoEta}, only the $\mathcal{B}(\eta\to\pi^+\pi^-\pi^0)$ is in agreement within one standard deviation. 
		The ratios of the measured BFs of $\eta$ are summarized in Table~\ref{table_sumRatio}, which are in agreement with CLEO’s result within two standard deviations. The sum of the four BFs, which provides a first constraint on the unknown decay modes of $\eta$, is $(99.24\pm 0.09\pm2.31)$\%, where the first error is statistical and the second systematic.

		\begin{table}[htbp]
		\begin{center}
			\caption{Comparison of ratios of BFs. The first error is statistical and the second systematic.}\label{table_sumRatio}
			\setlength{\tabcolsep}{0.2cm}{
			\begin{tabular}{lcc}\hline\hline
				& \multicolumn{2}{c}{$\mathcal{B}(\eta\to X)$/$\mathcal{B}(\eta\to\gamma\gamma)$}\\\cline{2-3}
				\multicolumn{1}{c}{$X$} & This Work & CLEO \\\hline
				$\pi^0\pi^0\pi^0$\color{white}{\Large{1}} & $\raCb\pm\raCbEsta\pm\raCbEsys$ & $0.884 \pm 0.022 \pm 0.019$ \\
				$\pi^+\pi^-\pi^0$ & $\raCd\pm\raCdEsta\pm\raCdEsys$ & $0.587 \pm 0.011 \pm 0.009$ \\
				$\pi^+\pi^-\gamma$ & $\raCc\pm\raCcEsta\pm\raCcEsys$ & $0.103 \pm 0.004 \pm 0.004$ \\
				\hline\hline
			\end{tabular}
			}
		\end{center}
		\end{table}

	\acknowledgements
The BESIII collaboration thanks the staff of BEPCII and the IHEP computing center for their strong support. This work is supported in part by National Key Research and Development Program of China under Contracts Nos. 2020YFA0406300, 2020YFA0406400; National Natural Science Foundation of China (NSFC) under Contracts Nos. 11625523, 11635010, 11735014, 11822506, 11835012, 11935015, 11935016, 11935018, 11961141012, 12022510, 12025502, 12035009, 12035013, 12061131003; the Chinese Academy of Sciences (CAS) Large-Scale Scientific Facility Program; Joint Large-Scale Scientific Facility Funds of the NSFC and CAS under Contracts Nos. U1732263, U1832207; CAS Key Research Program of Frontier Sciences under Contract No. QYZDJ-SSW-SLH040; 100 Talents Program of CAS; INPAC and Shanghai Key Laboratory for Particle Physics and Cosmology; ERC under Contract No. 758462; European Union Horizon 2020 research and innovation programme under Contract No. Marie Sklodowska-Curie grant agreement No 894790; German Research Foundation DFG under Contracts Nos. 443159800, Collaborative Research Center CRC 1044, FOR 2359, GRK 214; Istituto Nazionale di Fisica Nucleare, Italy; Ministry of Development of Turkey under Contract No. DPT2006K-120470; National Science and Technology fund; Olle Engkvist Foundation under Contract No. 200-0605; STFC (United Kingdom); The Knut and Alice Wallenberg Foundation (Sweden) under Contract No. 2016.0157; The Royal Society, UK under Contracts Nos. DH140054, DH160214; The Swedish Research Council; U. S. Department of Energy under Contracts Nos. DE-FG02-05ER41374, DE-SC-0012069



\begin{thebibliography}{99}
		\bibitem{etaRev2007} H. F. Jones, \emph{Groups, Representations and Physics} (Hilger, Bristol, England, 1990), p.150.
		\bibitem{etaRev2019} S. D. Bass and P. Moskal, Rev. Mod. Phys. {\bf 91}, 015003 (2019)
		\bibitem{IntroTwoGam} M. Poppe, Int. J. Mod. Phys. A {\bf 1} 545 (1986).
		\bibitem{IntroQuarkMass} A. Deandrea, A. Nehme and P. Talavera, Phys. Rev. D {\bf 78}, 034032 (2008).
		\bibitem{IntroChiralDynamic1} R. Escribano, P. Masjuan and J. J. Sanz-Cillero, J. High Energ. Phys. {\bf 2011}, 94 (2011).
		\bibitem{IntroChiralDynamic2} S. Scherer, Nucl. Phys. A {\bf 623}, 215 (1997).
		\bibitem{IntroQCDSymmetry} A.~Kup\'{s}\'{c} and A.~Wirzba, J. Phys. Conf. Ser. {\bf 335}, 012017 (2011).
		\bibitem{IntroBSM} R. Escribano and E. Royo, Eur. Phys. J. C {\bf 80}, 1190 (2020).
		\bibitem{bibGamEtaP} M. Ablikim \emph{et al.} [BESIII Collaboration], Phys. Rev. Lett. {\bf 122}, 142002 (2019).
		\bibitem{pdg2019} P.A. Zyla \emph{et al.} [Particle Data Group], Prog. Theor. Exp. Phys. {\bf 2020}, 083C01 (2020).
		\bibitem{cleoEta} A. Lopez \emph{et al.} [CLEO Collaboration], Phys. Rev. Lett. {\bf 99}, 122001 (2007).
		\bibitem{PCF} Z. R. Xu and K. L. He, Chin. Phys. C {\bf 36}, 742 (2012).
		\bibitem{Ablikim:2009aa} M.~Ablikim {\it et al.} [BESIII Collaboration], Nucl.\ Instrum.\ Meth.\ A {\bf 614}, 345 (2010).
		\bibitem{Yu:IPAC2016-TUYA01} C.~H.~Yu {\it et al.}, Proceedings of IPAC2016, Busan, Korea, 2016, doi:10.18429/JACoW-IPAC2016-TUYA01.
		\bibitem{Ablikim:2019hff} M.~Ablikim {\it et al.} [BESIII Collaboration], Chin. Phys. C {\bf 44}, 040001 (2020).
		\bibitem{etof} X.~Li {\it et al.}, Radiat. Detect. Technol. Methods {\bf 1}, 13 (2017); Y.~X.~Guo {\it et al.}, Radiat. Detect. Technol. Methods {\bf 1}, 15 (2017); P.~Cao {\it et al.}, Nucl.\ Instrum.\ Meth.\ A {\bf 953}, 163053 (2020).
		\bibitem{nJpsi0912} M. Ablikim \emph{et al.} [BESIII Collaboration], Chin. Phys. C {\bf 41}, 013001 (2017).
		\bibitem{geant4} S. Agostinelli \emph{et al.} [GEANT4 Collaboration], Nucl. Instrum. Methods Phys. Res., Sect. A {\bf 506}, 250 (2003).
		\bibitem{MCPackage} Z. Y. Deng \emph{et al.}, Chin. Phys. C {\bf 30}, 371 (2006).
		\bibitem{BESIIIDetectorShape} Y. T. Liang, B. Zhu, Z. Y. You \emph{et al.}, Nucl. Instrum. Meth. A {\bf 603}, 325 (2009).
		\bibitem{BESIIIDetectorShapeB} Z. Y. You, Y. T. Liang, Y. J. Mao, Chin. Phys. C {\bf 32}, 572 (2008).
		\bibitem{kkmc} S. Jadach, B. F. L. Ward and Z. Was, Phys. Chem. Comm. {\bf 130}, 260 (2000); S. Jadach, B. F. L. Ward and Z. Was, Phys. Rev. D {\bf 63}, 113009 (2001).
		\bibitem{evtgen} R. G. Ping, Chin. Phys. C {\bf 32}, 599 (2008); D. J. Lange, Nucl. Instrum. Methods Phys. Res., Sect. A {\bf 462}, 152 (2001).
		\bibitem{lundcharm} J. C. Chen, G. S. Huang, X. R. Qi, D. H. Zhang and Y. S. Zhu, Phys. Rev. D {\bf 62}, 034003 (2000); R. L. Yang, R. G. Ping and H. Chen, 
		Chin. Phys. Lett. {\bf 31}, 061301 (2014).
		\bibitem{photos} E.~Richter-Was, 
		Phys. Lett. B {\bf 303}, 163 (1993).
		\bibitem{evtgenGuide} A. Ryd, D. Lange \emph{et al.}, "EvtGen: A Monte Carlo Generator for B-Physics", EVTGEN-V00-11-07.
		\bibitem{etaTo3PiGen} M. Ablikim \emph{et al.} [BESIII Collaboration], Phys. Rev. D {\bf 92}, 012014 (2015).
		\bibitem{etaToGamLLGen} N. Qin \emph{et al.}, Chin. Phys. C {\bf 42}, 013001 (2018).
		\bibitem{babayaga1} C. M. Carloni Calame \emph{et al.}, Nucl. Phys. Proc. Suppl. {\bf 131}, 48 (2004).
		\bibitem{babayaga2} C. M. Carloni Calame \emph{et al.}, Phys. Lett. B {\bf 520}, 16 (2001).
		\bibitem{babayaga3} C. M. Carloni Calame \emph{et al.}, Nucl. Phys. B {\bf 584}, 459 (2000).
		\bibitem{eeEtaGen} L. M. Gu, H. B. Li, X. X. Ma and M. Z. Yang, Phys. Rev. D {\bf 100}, 016018 (2019).
		\bibitem{rhoPiGen} M. Ablikim \emph{et al.} [BESIII Collaboration], Phys. Lett. B {\bf 710} (2012).
		\bibitem{CBShape} M. Oreglia. PhD thesis, SLAC-R-236 (1980).
		\bibitem{eeEtaBF} M. Ablikim \emph{et al.} [BESIII Collaboration], Phys. Rev. D {\bf 99}, 012006 (2019).
		\bibitem{rhoPiBF2} D. Coffman \emph{et al.} [MARKIII Collaboration], Phys. Rev. D {\bf 40}, 3788 (1989).
		\bibitem{omegaEtaBF} M. Ablikim \emph{et al.} [BES Collaboration], Phys. Rev. D {\bf 73}, 052007 (2006).
		\bibitem{RooBukinLink2} A. D. Bukin, Fitting function for asymmetric peaks, arXiv:0711.4449.
		\bibitem{sysKmfitC0912} M. Ablikim \emph{et al.} [BESIII Collaboration], Phys. Rev. D {\bf 87}, 012002 (2013).
		\bibitem{sysKmfitN0912A} M. Ablikim \emph{et al.} [BESIII Collaboration], Phys. Rev. Lett. {\bf 104}, 132002 (2010).
		\bibitem{sysKmfitN0912B} M. Ablikim \emph{et al.} [BESIII Collaboration], Phys. Rev. D {\bf 94}, 072005 (2016).
	\end{thebibliography}
\end{document}